\documentclass[12pt,preprint]{aastex}
\usepackage{epsfig}
\usepackage{graphicx}
\usepackage{color}
\usepackage{enumerate}
\usepackage{ulem}

\newcommand{\msun}{{\rm M_\odot}}
\newcommand{\rsun}{{\rm r_\odot}}

\newcommand{\kms}{{\rm km~s^{-1}}}

\shorttitle{Interruption of Tidal Disruption Flares By SMBHBs}
\shortauthors{Liu et al.}

\begin{document}

\title{Interruption of Tidal Disruption Flares By
  Supermassive Black Hole Binaries}

\author{F. K. Liu\altaffilmark{1,2}, S. Li\altaffilmark{1}, and
  Xian Chen\altaffilmark{1,3}}

\altaffiltext{1}{Astronomy Department, Peking University, 100871
  Beijing, China;
  fkliu@bac.pku.edu.cn, lis@bac.pku.edu.cn, chenx@bac.pku.edu.cn}
\altaffiltext{2}{Kavli Institute for Astronomy and Astrophysics,
  Peking University, 100871 Beijing, China}
\altaffiltext{3}{Department of Astronomy, University of California,
  Santa Cruz, CA 96064, USA}

\begin{abstract}
Supermassive black hole binaries (SMBHBs) are products of  
galaxy mergers, and are important in testing $\Lambda$ cold dark
matter cosmology 
and locating gravitational-wave-radiation sources. A unique
electromagnetic signature of SMBHBs in galactic nuclei is essential in
identifying the binaries in observations from the IR band through optical
to X-ray. Recently, the flares in optical, UV, and X-ray caused
by supermassive black holes (SMBHs) tidally disrupting nearby stars 
have been successfully used to observationally probe
single SMBHs in normal galaxies. In this 
Letter, we investigate the accretion of the gaseous debris of a
tidally disrupted star by a SMBHB. Using both stability analysis
of three-body systems and numerical scattering experiments, 
we show that the accretion of stellar debris gas, which initially decays
with time $\propto t^{-5/3}$, would stop at a time $T_{\rm tr} \simeq
\eta T_{\rm b}$. Here, $\eta \sim0.25$ and $T_{\rm b}$ is the orbital
period of the SMBHB.  After a period of interruption, the
accretion recurs discretely at time $T_{\rm r} \simeq \xi T_b$,
where $\xi \sim 1$. Both $\eta$ and $\xi$ sensitively depend on the
orbital parameters of the tidally disrupted star at the tidal
radius and the orbit eccentricity of SMBHB. The interrupted
accretion of the stellar debris gas gives rise to an interrupted 
tidal flare, which could be used to identify SMBHBs in
non-active galaxies in the upcoming transient surveys. 
\end{abstract}


\section{Introduction}
\label{introduction}

Supermassive black hole binaries (SMBHBs) are predicted by the
hierarchical galaxy formation model in $\Lambda$ cold dark
matter ($\Lambda$CDM) cosmology \citep{beg80,vol03}. After a
SMBHB at the center of merging systems
become hard, it would stall at the hard radius for a timescale
even longer than the Hubble time if spherical two-body relaxation
dominates \citep{beg80}. However, recent investigations suggested that
the hardening rates of SMBHBs can be boosted and SMBHBs may 
coalesce within a Hubble time either due to various stellar dynamical
processes other than the spherical two-body relaxation
\citep{yu02,cha03,mer04,ber06,ses08}, or due to gas dynamics
\citep[][and references therein]{gou00,liu03,cop09}.

The strong gravitational wave (GW) radiation generated by coalescing
supermassive black holes (SMBHs) is the main target of the GW detector
the Laser Interferometer Space Antenna (LISA) and of the Pulsar Timing
Array (PTA) program. Because of the poor accuracy of both LISA and PTA in
locating GW radiation sources, it is of key importance to
detect electromagnetic counterparts (EMCs) of GW radiation
sources. Identifying SMBHBs by their EMCs is also
essential to constraining the poorly understood galaxy-merger
history. Several EMCs have been suggested in the literature to probe
SMBHBs and their coalescence: (1) precession of jet orientation
and its acceleration in radio galaxies during the in-spiraling of
SMBHBs \citep{beg80,liu07}, (2) optical periodic outbursts
in blazars due to the interaction between SMBHB and accretion
disk \citep{sil88,liu95,liu06,liu02,val08,hai09}, (3)
jet reorientation in X-shaped radio galaxies due to the exchange of
angular momentum between SMBHB and accretion disk \citep{liu04}, (4)
two systems of broad emission lines (BELs) in quasars \citep{bor09},
(5) intermittent activity in double-double radio galaxies at
binary coalescence \citep{liu03}, (6) X-ray, UV, optical, and IR
afterglow following binary coalescence
\citep{mil05,shi08a,lip08,sch08a}, and (7) systematically shifted BELs
relative to narrow emission lines \citep{mer06,kom08} and off-center
active galactic nuclei \citep[AGNs;][]{mad04,loe07} because of SMBH
GW radiation recoil.

All the above observational signatures require gas accretion disks
around SMBHBs. The SMBHBs in gas-poor 
galactic nuclei are difficult to detect, because the SMBHs are
dormant. However, a dormant SMBH could be temporarily activated
by tidally disrupting a star passing by and accreting the
disrupted stellar debris \citep{hil75}. A tidal flare
decays typically as a power law $t^{-5/3}$ \citep{ree88,phi89}, which
has been observed in several non-active galaxies
\citep{kom99,kom02,hal04,esq08,gez08,gez09}. Recently, \citet{che08} 
and \citet{che09} calculated the tidal
disruption rate in SMBHB systems at different evolutionary stages,
and found that it is significantly different from the typical rate 
for single SMBHs by orders of magnitude. This great difference
of the flaring rates enables one to statistically constrain the SMBHB
population in normal galaxies \citep{che08}, but identifying SMBHBs
individually is still difficult at present. In this 
Letter, we investigate the influence of a SMBHB on the accretion of
tidally disrupted stellar plasma and on the tidal flare. We show that
the accretion of the stellar debris is interrupted by the SMBHB and
this interruption can be taken as a key distinguishable observational
signature for SMBHBs in gas-poor galactic nuclei. 

\section{Tidal Disruption in a SMBHB system}
\label{anal}

A star with mass $m_*$ and radius $r_*$ will be tidally disrupted by a
SMBH with mass $M_\bullet$ if it approaches 
the black hole (BH) within the tidal radius
\begin{equation}
  r_t \simeq \mu r_{*}(M_\bullet/m_{*})^{1/3}
  \label{rt}
\end{equation}
\citep{hil75,ree88,phi89},
where $\mu$ is a dimensionless parameter of order unity. We focus
on tidal disruptions of solar type stars with solar radius
$\rsun$ and solar mass $\msun$ by SMBHs with $M_\bullet \la 10^8
\msun$, which are detected in the tidal-flare surveys
\citep{kom99,kom02,esq08,gez08,gez09}. For more massive BHs, the tidal
radii become smaller than the event horizon \citep{iva06}. After tidal
disruption at the tidal radius,
the specific energy $E$ across the star ranges from $-E_b-\Delta E$ to
$-E_b+\Delta E$, where $E_b$ is the orbital binding energy of the
star, $ \Delta E  =  kGM_\bullet r_*/r_t^2$ is the spread in specific
energy across the stellar radius, and $k=1$ if the tidal spin-up of
the star is negligible or $k=3$ if the star is spun-up to the break-up
angular velocity \citep{ree88,lac82,li02}. Since $\Delta E$
is $(M_\bullet/M_*)^{1/3}$ times larger than the internal binding
energy of the star and the interaction between the stellar debris
 is negligible after the tidal disruption \citep{eva89}, each fluid
 element moves ballistically.

If we neglect the influence of the companion SMBH, the bound debris
with $E<0$ should move in Keplerian orbits with
eccentricity of about unity and fall back to the tidal radius
after one Keplerian period, $T=2\pi GM_\bullet/(-2E)^{3/2}$.
The debris continuously returning to the tidal radius generates a
 falling-back rate $\dot{m}= (dm/dE) (dE/dt)$. For Keplerian orbits, 
 $dE/dt=dE/dT\propto t^{-5/3}$. If we assume $dm/dE$ following a
 constant distribution in the energy range $-E_b-\Delta E$ to
 $-E_b+\Delta E$ \citep[see e.g.][]{ree88,eva89,lod08}, the
 falling-back rate evolves as
\begin{eqnarray}
  \dot{m} & \simeq & {m_{*} \over 3 T_{\rm min}} \left({t - T_D
      \over T_{\rm min}}\right)^{-5/3}
  \label{Mdot}
\end{eqnarray}
for $t \ge T_{\rm D} + T_{\rm min}$, where $T_D$ is the disruption
time and
\begin{eqnarray}
  T_{\rm min} & \simeq& 2 \pi GM_\bullet \left( 2\Delta E\right)^{-3/2}
  \nonumber \\
  & \simeq & 0.355 \, {\rm yr} \, k^{-3/2} \left({R_* \over
  \rsun}\right)^{3/2} \left({m_{*} \over \msun}\right)^{-1}
  \left({M_\bullet \over 10^7\msun}\right)^{1/2} ,
\label{eq:tmin}
\end{eqnarray}
is the returning time for the most bound debris (typically
$|E_b|\ll\Delta E$).  It is assumed that once the bound material comes
back to the tidal disruption radius, it loses
kinetic energy due to strong shocks because of the interaction
  between fluid elements and is circularized on a timescale much
shorter than $T$ to form an orbiting torus at a radius about
$r_{\rm c} = 2 r_t$ around the SMBH
\citep[e.g.][]{ree88,phi89,ulm99,li02}. Both the
radiative dissipation of the shock energy and the accretion of the gas
torus onto the BH will give rise to an X-ray/UV
flare, decaying with $(t-t_{\rm D})^{-5/3}$.

For a SMBHB system, the stellar-disrupting BH could be either the
primary (with mass $M_1$) or the secondary (with 
mass $M_2$). Since the probability of stellar disruption by the
secondary is relatively low if the two BHs are 
very unequal  \citep{che08,che09}, here we assume the primary BH to be
the stellar-disrupting one. Because the 
interaction between the fluid elements is negligible before
  they come back to the tidal disruption radius, the SMBHB and each
bound fluid element constitute a restricted three-body system. When
the orbit of a bound fluid element is inside the SMBHB orbit, the
system is called S-type. An S-type three-body system could
be stable for a long time only when the system is hierarchical, namely,
the system consists of an inner binary (the bound fluid element and
the primary BH) on a nearly Keplerian orbit with semi-major
axis $a_{\rm f}$ and eccentricity of near unit, and an outer
binary with eccentricity $e$ and semi-major axis $a_{\rm b} \gg a_{\rm
f}$ in which the secondary BH orbits the mass center of the inner binary.
The orbital change of a stable S-type system is negligible on the
fluid-element dynamical timescale, so the falling-back
stellar debris may interact with one another at the tidal disruption
radius and finally be accreted with accretion rate
given in Equation~(\ref{Mdot}). However, for those less-bound fluid
elements with $a_{\rm f}$ larger than a critical radius $a_{\rm cr}$,
the orbit of the triple
system becomes chaotic. For a triple system with an angle $\theta$
between the angular momenta of the inner and outer binaries, $a_{\rm
  cr}$ can be semiempirically given by
\begin{eqnarray}
  a_{\rm b} / a_{\rm cr} & = & 2.8 \left(1 + q \right)^{2/5} \left(1 +
  e \right)^{2/5} \nonumber \\
  &&\times \left(1 - e \right)^{-6/5} \left(1 - 0.3
  \theta/180^{\circ} \right)
  \label{stab_condition}
\end{eqnarray}
\citep{mar01}, where $q=M_2/M_1<1$ is the BH mass ratio.

Because of the nonlinear overlap of the multiple resonances
\citep{mar07} in the chaotic triple systems, the fluid 
elements significantly exchange angular momentum and energy 
with the SMBHB and change their orbit 
dramatically on the dynamical timescale of the triple system,
therefore would not return to the tidal radius to fuel the accreting
torus and to form continuous accretion. Although a fraction of
the fluid elements with chaotic orbits may return to the tidal
radius on a timescale much longer than its Keplerian period, others 
would escape the system during the three-body interactions
(see our numerical simulations in Section~\ref{methods}). Therefore, 
Equation~(\ref{stab_condition}) implies that the continuous
accretion of tidally disrupted stellar plasma stops with the
accretion of the last fluid element with $a<a_{cr}$, this occurring at
a time 
\begin{eqnarray}
  T_{\rm tr}&\approx& {T_{\rm b} \over 4.7} \left({1 +
    q}\right)^{-1/10} \left({1 + e}\right)^{-3/5}
  \nonumber \\
  & & \times \left(1 - e\right)^{9/5} \left(1 - 0.3
  \theta/180^{\circ} \right)^{-3/2} ,
  \label{ttrun}
\end{eqnarray}
where $T_{\rm b}$ is the SMBHB orbital period. The eccentricity of
hard SMBHBs is moderate \citep{mil01} and minor mergers with $q\ll1$
are the most common in the hierarchical galaxy formation model
\citep{vol03}. We re-write Equation~(\ref{ttrun}) as $T_{\rm tr}=\eta
T_b$, where $\eta$ is in the range $(0.21,0.36)$ for $e \sim 0$ and
$q\ll 1$, and independent of $a_{\rm b}$. For a SMBHB with semi-major
axis $a_{\rm b} = a_h/\beta$ residing in a stellar cluster with
velocity dispersion $\sigma_*$, where
 \begin{equation}
 a_h=\frac{GM_1 M_2} {4 (M_1 + M_2)\sigma_*^2}
 \label{ah}
 \end{equation}
is the hard radius and $\beta$ is the hardness, the accretion is
interrupted at
\begin{equation}
  T_{\rm tr} = \eta T_{\rm b} \simeq 6.2 \, {\rm yr} \sigma_{110}^{-3}
  M_7 q_{-1}^{3/2} (1+q)^{-2} \left(\eta\over0.25\right) \left(\beta
  \over 10 \right)^{-3/2} ,
 \label{tr_th}
\end{equation}
where $\sigma_{110} = \sigma_* / 110~\kms$, $M_7 = M_1 / 10^7~\msun$
and $q_{-1} = q/0.1$.

\section{Numerical simulations and results}
\label{methods}

Because the fluid elements of a tidally disrupted star move like test
particles before falling back to the tidal radius, their
evolution could be correctly simulated with scattering
experiments. We use $N=10^6$ particles to resolve the stellar debris
of a disrupted star and integrate their trajectories in restricted
three-body systems. The particles are logarithmically sampled in
the binding-energy range $[E_{\rm thd},\Delta E]$,
where $E_{\rm thd}=G(M_1+M_2)/2.5a_b$ is a binding energy
corresponding to an orbital period of $4T_b$ and $\Delta E=GM_1
r_*/r_t^2$ is the spread in specific energy for a non-spinning
star. For a typical tidal-disruption event with
$|E_b|\ll\Delta E$, the $i$th particle has a binding energy
$E_i=\Delta E(E_{\rm thd}/\Delta E)^{(i-0.5)/N}$, pericenter
velocity $V_i=(2E_{i}+2GM_1/r_t)^{1/2}$, and mass $m_i=0.5m_*\Delta
E_i/\Delta E$, where $\Delta E_i=E_i\ln(0.5E_{\rm thd}/\Delta E)/N$
is the size of the $i$th energy bin.

The evolution of the three-body systems is computed in a frame
centered on  the mass center of the SMBHB with $X$--$Y$ plane aligned
with the SMBHB orbital plane. We assume $e=0$ in this work for
simplicity, and that the secondary BH initially lies on the positive
$X$ axis and moves in the direction of positive $Y$.
In each experiment, a test particle starts with velocity
$V_i$ from the pericenter at tidal radius $r_t$ about the
primary BH, and the orbit is determined by three initial parameters
(determined by the orbital parameters of the disrupted star at
tidal radius) (1) the inclination angle $\theta $ between the orbital
planes of the particle and the SMBHB, (2) the longitude of ascending
node $\Omega$, and (3) the argument of pericenter $\omega$. With
the initial parameters, we integrate the equations of
motion in the pseudo-Newtonian potential
$\phi = GM_{\rm \bullet} /(r - r_{\rm g})$ \citep{pac80},
where $r_{\rm g}$ is the Schwarzschild radius of BHs, using
an explicit Runge-Kutta method of order 8 \citep{hai87}. Stellar
debris which falls back to within $2r_t$ from the
primary BH is assumed to be circularized via shocks and be accreted
instantaneously. Therefore, if a particle with id $i$ passes by the
primary BH within a distance $r_{\rm acc}=2 r_t$ \footnote{We test our
simulations with $r_{\rm acc} = 1.5 r_t$, $3r_t$, and $6r_t$ and the
results are nearly independent of $r_{\rm acc}$.}, the integration is
stopped and the time $t_i$ is recorded. Otherwise, the integration
continues until time $4T_b$. The accretion rate of
the stellar debris is calculated using the recorded $m_i$ and $t_i$.

\begin{figure}
\epsscale{1.0} \plotone{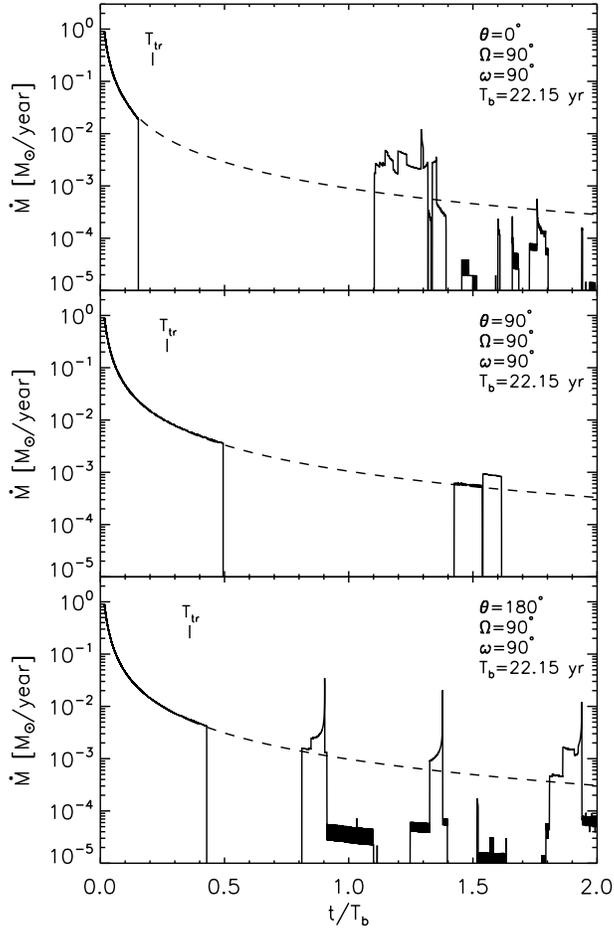}
\caption{Accretion rates of gaseous debris in units of solar
  mass per year vs. time. Time starts at tidal disruption
  and is in unit of the binary orbital period $T_{\rm
  b} = 22.15\, {\rm yr}$. The solid lines are the simulation results
  and the dashed lines give the standard $\sim t^{-5/3}$
  in single BH systems. $T_{\rm tr}$ marks the interruption time
  estimated with Equation~(\ref{ttrun}).  The upper, middle, and bottom
  panels are, respectively, for $\theta=0\degr$, $90\degr$, and
  $180\degr$.
  \label{fiducial}}
\end{figure}

Figure~\ref{fiducial} shows the accretion rate of stellar
debris in our fiducial simulations with $q=0.1$, $M_7=1$,
$\beta=10$, $\sigma_{110}=0.9545$, $r_*=\rsun$, $m_*=\msun$,
$\Omega=90\degr$, and $\omega=90\degr$. Our results suggest that the
interruption of accretion occurs at $(0.15-0.5)T_b$,
depending on $\theta$. In Figure~\ref{fiducial}, a better agreement
between the numerical value $t_{\rm tr}$ and the analytical estimate
$T_{\rm tr}$ for $\theta=0\degr$ and $180\degr$ is by coincidence,
because $\eta$ not only depends on $\theta$ but also on $\Omega$ and
$\omega$. To illustrate this, we did 100 numerical simulations
with random $\cos(\theta)$, $\Omega$, and $\omega$. The results show that
$\eta$ ranges from $0.15$ to $0.5$ with a mean value $0.25$,
consistent with the analytical mean value 0.27 given by
Equation~(\ref{ttrun}). Figure~\ref{fiducial} also shows that after
being interrupted for about $(1 - \eta)T_b$, accretion
recurs and ``accretion islands'' emerge discretely.
The accretion rate at the islands is variable and can be
larger than the corresponding value for single BH. The
duration of interruption and the periods of the accretion
islands decrease with $\theta$, the shortest occurring at $\theta=
180\degr$.

\begin{figure}
\epsscale{1.0} \plotone{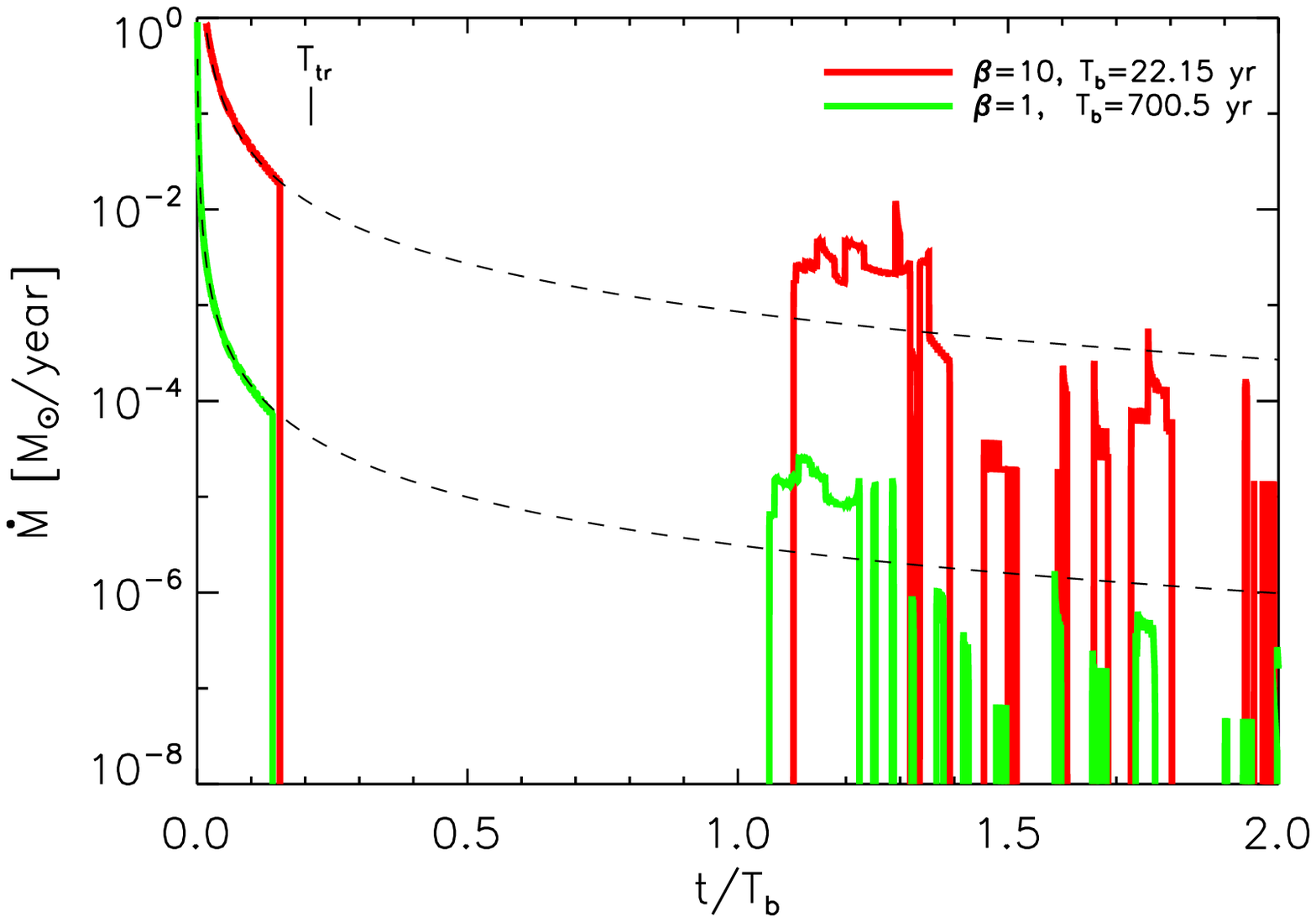} \caption{Accretion rates of
  gaseous debris vs. time for different $\beta$
  at $\theta=0$. The red and green solid lines correspond to
  $\beta=10$ and $\beta=1$, respectively. The other parameters are the
  same as those in Figure~\ref{fiducial}.\label{bdep}}
\end{figure}

To investigate the effect of different $a_b$ on the results,
we ran a scattering experiment for $\beta=1$ and $\theta=0$, keeping
the other parameters as in Figure~\ref{fiducial}.
 The result is given in Figure~\ref{bdep}. Figure~\ref{bdep} shows
 that the interruption occurs at a slightly earlier time $t_{\rm
   tr}/T_b$ as hardness
 $\beta$ decreases, which seems inconsistent with the analytical
 prediction given by Equation~(\ref{ttrun}).  However, we should note
 that Equation~(\ref{ttrun}) is obtained statistically by averaging
 over different $\Omega$ and $\omega$. When we did the simulations with
 different $\Omega$ and $\omega$, the averaged interruption time for
 different $\beta$ resides in the predicted range. Accordingly, in the
 scattering experiments with different BH mass $M_1$ ($M_7 = 0.1 $) and
 mass ratios $q$ ($q =1, 0.02$), we find that $\eta$ weakly depends on
 $q$ and $M_1$. Our numerical results are consistent with
 the analytical estimate given by Equation~(\ref{ttrun}), implying
 that the dependence of $\eta$ on $\beta$, $q$, and $M_1$ is much
 weaker than on $\theta$, $\Omega$, and $\omega$.

\section{Discussions}
\label{discussion}

We investigated the accretion of tidally disrupted stellar debris in
SMBHB systems. For simplicity, we assumed for the tidal debris gas:
(1) a constant distribution of 
mass in binding energy at the tidal radius, (2) ballistic motion and
negligible interaction of the fluid elements after tidal disruption,
and (3) instant circularization of the debris gas probably due to
shocks because 
of interaction between the fluid elements returned to within a radius
two times the tidal radius. The first two assumptions have been 
justified by numerical hydrodynamic simulations
\citep[e.g.][]{eva89,lod08}, but the third one needs to be 
verified by hydrodynamic simulations capable of capturing strong
shocks. With these three assumptions, the fluid 
elements and SMBHB compose restricted three-body systems, so we can
investigate the accretion of the gaseous 
stellar debris by analyzing the stability of the three-body systems
using the {\it resonance overlap stability criterion}
\citep{mar07}. Because of the chaotic nature induced by
the nonlinear overlap of several orbital mean 
motion resonances, the fluid elements inside the chaotic regions
significantly change their orbits on a dynamical timescale and do
not return to the tidal radius to fuel the BH, leading to the
interruption of accretion. We also investigated the evolution of the 
stellar debris using three-body scattering experiments. Our results
obtained both analytically and numerically show that the accretion
rate of the debris gas decreases with a power law $\dot{M} \sim
t^{-5/3}$ until a critical 
time $T_{\rm tr}$. At time $t>T_{\rm tr}$, the
  accretion pauses until about one SMBHB orbital period
$T_{\rm b}$. $T_{\rm tr}$ relates to $T_{\rm b}$ with $T_{\rm tr} =
\eta T_{\rm b}$, where $\eta$ is typically $0.25$ and in the range
$0.15 \la \eta \la 0.5$, depending on the initial orbital parameters
($\theta, \Omega, \omega$) of the fluid elements at tidal radius and
on the eccentricity of the SMBHB, but being nearly independent of the
SMBHB semimajor $a_{\rm b}$, BH masses, and mass ratio
$q$. This suspension of accretion would result in an
interruption of the tidal flare, although the residual accretion disk
may still radiate weak optical, UV, and X-ray emission during the
interruption. 

Our numerical results indicate that the accretion of the gaseous
stellar debris restarts at a time $T_{\rm  r}$,  leading to a
  flicker of flare. The exact recurring time, $T_{\rm r} = \xi T_{\rm
  b}$, depends on the initial orbital parameters $\theta$, $\Omega$,
and $\omega$, but our numerical results suggest that $\xi$ is of order
unity and $1 \la \xi \la 1.5$. The interruption 
timescale of the tidal flare in a SMBHB system is
\begin{equation}
  \Delta{T} = T_{\rm r} - T_{\rm tr} \simeq (\xi - \eta) T_{\rm b} .
\label{int_timescale}
\end{equation}
Our numerical simulations suggest that the accreted plasma during the
discrete accretion consists of 
both the tidal debris gas falling back to the tidal radius after one 
Keplerian time and a fraction of those fluid elements with
chaotic orbits falling back to the accreting torus on a timescale 
longer than its Keplerian time. When $T_{\rm tr}$ and $\Delta{T}$ is
determined observationally, constraints on $\eta$ and $T_{\rm b}$
could be made if we take $\xi \simeq 1$.

In our simulations, we assume that the orbital binding energy
$E_b$ of the tidally disrupted star is negligible compared to the spread
in specific energy $\Delta E$. For such tidal-disruption events,
Equations~(\ref{eq:tmin}) and (\ref{tr_th}) imply that the standard
power-law decay and the interruption of tidal flares are detectable if 
\begin{eqnarray}
  \beta < \beta_{\rm max} &\approx& 130 \sigma_{110}^{-2} M_7^{1/3}
  q_{-1} (1+q)^{-4/3} \left({k\over2}\right)
\nonumber\\
  && \left(\eta\over0.25\right)^{2/3} \left({R_* \over
  \rsun}\right)^{-1} \left({m_{*} \over \msun}\right)^{2/3} .
\label{beta_lim}
\end{eqnarray}
If the tidally disrupted star is initially very bound to one of the
binary BHs so that $|E_b|$ is comparable to 
$\Delta E$ \citep{che09}, the time $T_{\rm min}$ would be much shorter
and the interruption of tidal flare is detectable even in an
ultra-hard SMBHB with $\beta\gg\beta_{\rm max}$. For a SMBHB with
$\beta \sim 100$, the semi-major axis of the binary $a_{\rm b}$ is
about $a_{\rm b} \simeq 9.3 \times 10^{2} r_{\rm g} q_{-1} (1+q)^{-1}
\sigma_{110}^{-2}$. The GW radiation emitted by such SMBHBs could be
detected by PTA. 

An interrupted tidal flare could be caught if $T_{\rm tr}$ is shorter
 than the mission duration of an transient survey 
 $t_{\rm sv} \sim 1-10 \, {\rm yr}$, which corresponds
to SMBHBs with $\beta \sim (5 - 100) \times 
q_{-1} M_7^{1/6}$. Here we use the $M_\bullet$--$\sigma_*$ relation 
\citep{tre02}. Because a hard SMBHB with 
$q=0.1$ spends most of its life time at $\beta\sim 5-100$
\citep{yu02,ses08}, we have a good chance to detect them 
with upcoming transient surveys. However, if one wants to catch an
interrupted tidal flare from SMBHBs emitting 
strong GW radiation ($\beta\sim 1000$), the time resolution of the
survey should be $\la 0.1 \, {\rm yr}$. Because 
a SMBHB spends a small fraction of its lifetime at this stage,
high-sensitivity and deep transient surveys are 
needed to accumulate many more tidal-disruption events.

\acknowledgments

We are grateful to X.-B. Wu, D.N.C. Lin, S. Aarseth, M. Valtonen
and S. Komossa for useful discussions.  Many thanks are due to
the referee, R.A. Mardling, for valuable comments. Thanks are
given to S. Justham for careful read of the manuscript. This work is
supported by NSFC (10573001) and Chinese
National 973 program (2007CB815405). XC thanks the China
Scholarship Council for financial support. The numerical computation
was carried on the SGI Altix 330 system at the Astronomy
Department of PKU.

\end{document}